\documentclass[twocolumn,aps,prapplied]{revtex4-1}

\usepackage{amsfonts}
\usepackage{amsmath}
\usepackage{graphicx}
\usepackage{bm}
\usepackage{amssymb}
\usepackage{dcolumn}
\usepackage{color}
\usepackage{multirow}
\usepackage{booktabs}
\usepackage[colorlinks,
linkcolor=blue,
anchorcolor=blue,
citecolor=blue,
urlcolor=blue,
]{hyperref}
\begin{document}
\title{Strain-Induced Room-Temperature Ferromagnetic Semiconductors with Large Anomalous Hall Conductivity in Two-Dimensional Cr$_2$Ge$_2$Se$_6$}

\author{Xue-Juan Dong$^1$, Jing-Yang You$^1$}

\author{Bo Gu$^{2,3}$}

\email[]{gubo@ucas.ac.cn}

\author{Gang Su$^{1,2,3}$}

\email[]{gsu@ucas.ac.cn}

\affiliation{$^1$ School of Physical Sciences, University of Chinese Academy of Sciences, Beijing 100049, China \\
$^2$ Kavli Institute for Theoretical Sciences,and CAS Center for Excellence in Topological Quantum Computation, University of Chinese Academy of Sciences, Beijing 100190, China \\
$^3$ Physical Science Laboratory, Huairou National Comprehensive Science Center, Beijing 101400, China}
     												
\date{\today}
\begin{abstract}
By density functional theory calculations, we predict a stable two-dimensional (2D) ferromagnetic semiconductor Cr$_2$Ge$_2$Se$_6$, where the Curie temperature $T$$_c$ can be dramatically enhanced beyond room temperature by applying a few percent strain. In addition, the anomalous Hall conductivity in 2D Cr$_2$Ge$_2$Se$_6$ and Cr$_2$Ge$_2$Te$_6$ is predicted to be comparable to that in ferromagnetic metals of Fe and Ni, and is an order of magnitude larger than that in diluted magnetic semiconductor Ga(Mn,As). Based on superexchange interactions, the enhanced $T$$_c$ in 2D Cr$_2$Ge$_2$Se$_6$ by strain can be understood by the decreased energy difference between 3$d$ orbitals of Cr and 4$p$ orbitals of Se. Our finding highlights the microscopic mechanism to obtain the room temperature ferromagnetic semiconductors by strain.
\end{abstract}

\maketitle

\section{Introduction}
Combining magnetism and semiconductor enables the development of magnetic semiconductors, a promising way to realize spintronic applications based on use of both charge and spin degrees of freedom in electronic devices \cite{Ohno1998,Dietl2010}. The highest Curie temperature of the most extensively studied magnetic semiconductor (Ga,Mn)As has been $T$$_c$ = 200 K \cite{Chen2011}, still far below room temperature. The room temperature ferromagnetic semiconductors are highly required by applications. Recent advances in magnetism in two-dimensional (2D) van der Waals materials have provided a new platform for the study of magnetic semiconductors \cite{Burch2018}. The Ising ferromagnetism with out-of-plane magnetization was observed in monolayer CrI$_3$ in experiment with $T$$_c$ = 45 K \cite{Huang2017}. The Heisenberg ferromagnetic state was obtained in 2D Cr$_2$Ge$_2$Te$_6$ in experiment with $T$$_c$ = 28 K in bilayer Cr$_2$Ge$_2$Te$_6$ \cite{Gong2017}, where the corresponding bulk was known as a layered ferromagnet with spin along $c$ axis and $T$$_c$ = 61 K in experiment \cite{Carteaux1995}. A large remanent magnetization with out-of-plane magnetic anisotropy was recently reported in experiment in the 6-nm film of Cr$_2$Ge$_2$Te$_6$ on a topological insulator (Bi,Sb)$_2$Te$_3$ with $T$$_c$ = 80 K \cite{Mogi2018}. The magnetic structure in monolayer CrI$_3$ and Cr$_2$Ge$_2$Te$_6$ was recently discussed in terms of Kitaev interaction \cite{Xu2018}. Photoluminescence and magneto-optical effects have also been discussed \cite{2018nature-physics-CrI3,2018PRBCrGeTe3}. The ferromagnetism with high $T$$_c$ was also reported in experiment in the monolayer VSe$_2$ \cite{Bonilla2018} and MnSe$_2$ \cite{OHara2018}. Despite the relatively small number of monolayer ferromagnetic materials realized in experiment, predicting promising candidates by first principles calculations can provide reliable reference for experiments  \cite{Shabbir2018,Sivadas2015}. Researchers have also studied possible 2D ferromagnetic materials by machine learning \cite{arxiv-machine-learning} and high-throughput calculations \cite{Liu2018-mean-field}.

Several methods are used to control the magnetic states in these recently discovered 2D materials. By tuning gate voltage, the switching between antiferromagnetic and ferromagnetic states in bilayer CrI$_3$ was obtained in experiment \cite{Huang2018}. Using gate voltage, the enhancement of ferromagnetism in 2D Fe$_3$GeTe$_2$ was observed in recent experiment \cite{Deng2018}. In 2D transition metal dichalcogenides, which are non-magnetic, the strain is used to modify the optical and electronics properties  \cite{Roldan2015}. The effect of strain is studied to affect the ferromagnetism in monolayer Cr$_2$Ge$_2$Te$_6$ \cite{Li2014-MonteCarlo} and CrX$_3$ (X = Cl, Br, I) \cite{Liu2016-MonteCarlo}. The electric field effect was discussed in experiment in magnetic multilayer Cr$_2$Ge$_2$Te$_6$ \cite{Xing2017}. Recently, heterostructures of these 2D materials have also been studied \cite{2017sciadvWSe2CrI3,Han-CrGeTe3-2018,2019nanoletterCrGeTe-Pt}.

In this paper, by density functional theory calculations we predict a stable 2D ferromagnetic semiconductor Cr$_2$Ge$_2$Se$_6$. We find that $T$$_c$ = 144 K in Cr$_2$Ge$_2$Se$_6$, which can be enhanced to $T$$_c$ = 326 K by applying 3$\%$ strain, and $T$$_c$ = 421 K by 5$\%$ strain. On the other hand, $T$$_c$ in 2D semiconductor Cr$_2$Ge$_2$Te$_6$ is about 30 K in our calculation, close to the value of $T$$_c$ = 28 K in recent experiment \cite{Gong2017}. In addition, the anomalous Hall conductivity in 2D Cr$_2$Ge$_2$Se$_6$ and Cr$_2$Ge$_2$Te$_6$ is predicted to be comparable to that in ferromagnetic metals of Fe and Ni \cite{Yao2004,Wang2006,Wang2007}, and is an order of magnitude larger than that in diluted magnetic semiconductor Ga(Mn,As) \cite{Jungwirth2003,Sinova2002}. The strain is found to decrease the energy difference between 3$d$ orbitals of Cr and 4$p$ orbitals of Se, which induces the enhanced ferromagnetic coupling based on the superexchange picture. Our finding highlights the microscopic mechanism to obtain the room temperature magnetic semiconductors by strain.

\section{Calculation Methods}
The density functional theory (DFT) calculations are done by the Vienna $ab$ $initio$ simulation package (VASP) \cite{Kresse1996}. The spin-polarized calculation with projector augmented wave (PAW) method, and general gradient approximations (GGA) in the Perdew-Burke-Ernzerhof (PBE)  exchange correlation functional are used. The spin-orbit coupling (SOC) is included in the calculations. Total energies are obtained with k-grids 9 $\times$ 9 $\times$ 1 by Mohnkhorst-Pack approach. The lattice constants and atom coordinates are optimized with the energy convergence less than 10$^{-6}$ eV and the force less than 0.01 eV/$\AA$, where a large vacuum of 20 $\AA$ is used to model a 2D system. The phonon calculations are performed with density functional perturbation theory (DFPT) by the PHONOPY code \cite{Togo2015-phono}. Size of the supercell is 3 $\times$ 3 $\times$ 1, where the displacement is taken by 0.01 $\AA$.

For the on-site Coulomb interaction U of Cr ion, the values in range of 3 $\sim$ 5 eV are usually reasonable for 3$d$ transition metal insulators \cite{Maekawa2004}, while a small value of U {\textless} 2 eV was adopted in the DFT calculation of 2D Cr$_2$Ge$_2$Te$_6$ \cite{Gong2017}. We fixed the parameter U = 4 eV in most of our following calculations for 2D Cr$_2$Ge$_2$Te$_6$ and Cr$_2$Ge$_2$Se$_6$, and will discuss the effect of different U parameters later.

Based on the DFT results, the Curie temperature is calculated by using the Monte Carlo simulations\cite{Liu2016-MonteCarlo,Li2014-MonteCarlo} based on the 2D Ising model, where a 60 $\times$ 60 supercell is adopted, and 10$^5$ steps are performed for every temperature to acquire the equilibrium. The anomalous Hall conductivity is calculated with Wannier90 code \cite{Mostofi2014-wannier} and WannierTools code \cite{Wuquansheng2018}.

\begin{figure}[tbp]
\includegraphics[width=8.5cm]{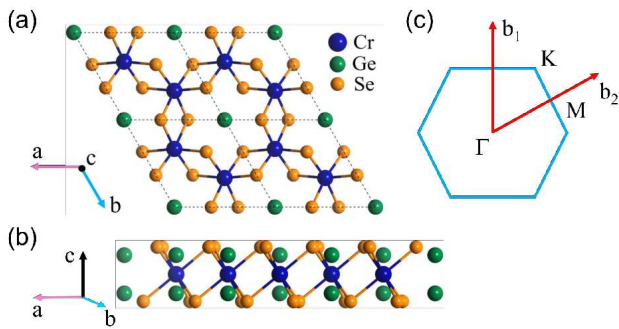}
\caption{ Crystal structure of Cr$_2$Ge$_2$Se$_6$ of (a) top view in a-b plane and (b) side view in a-c plane.
The two-dimensional Brillouin zone is shown in (c).}
\label{F-1}
\end{figure}

\begin{figure}[tbp]
\includegraphics[width=8.5cm]{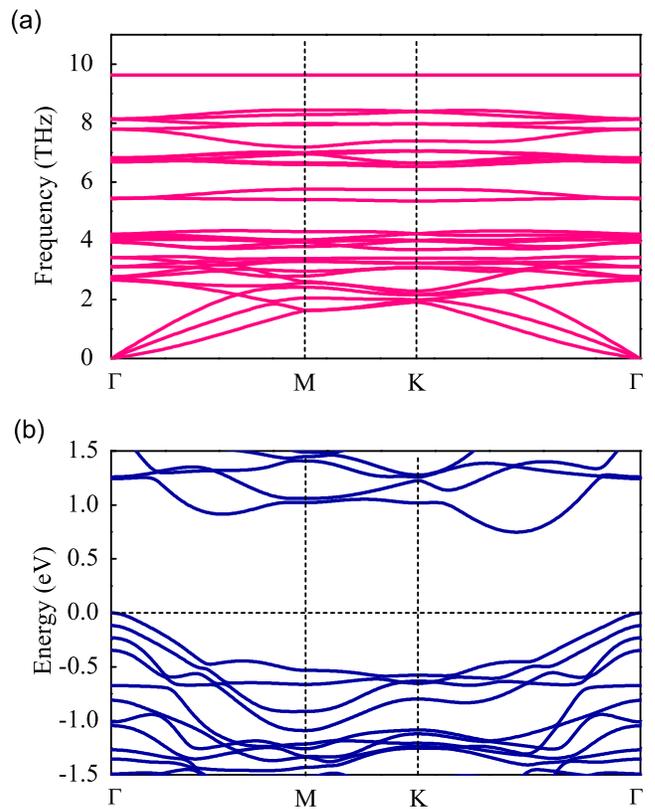}
\caption{(a) Phonon spectrum and (b) electron band structure of two-dimensional Cr$_2$Ge$_2$Se$_6$,
obtained by the spin-polarized GGA + SOC + U calculations.}
\label{F-2}
\end{figure}

\section{Crystal Stability}
We study the stability of the 2D new material Cr$_2$Ge$_2$Se$_6$. We propose this material guided by the recently reported 2D material Cr$_2$Ge$_2$Te$_6$ in experiment, and we replace Te by Se. The crystal structure of Cr$_2$Ge$_2$Se$_6$ is shown in Fig. \ref{F-1} (a) for top view and \ref{F-1} (b) for side view, where the space group number is 162 (P-31m). The structure is obtained by the fully relaxed calculation. The optimized lattice constant of 2D Cr$_2$Ge$_2$Se$_6$ is calculated as 6.413 $\AA$, which is smaller than the lattice constant 6.8275 $\AA$ of Cr$_2$Ge$_2$Te$_6$ in experiment. It is reasonable because Se has a smaller radius than Te. We examine the stability in terms of phonon spectrum, where the Brillouin zone is shown in Fig. \ref{F-1}(c). As shown in Fig. \ref{F-2}(a), there is no imaginary frequency in the phonon dispersion near $\Gamma$ point. It predicts that the crystal structure of monolayer Cr$_2$Ge$_2$Se$_6$ is stable.

 To further check the crystal stability, we study the formation energy of monolayer Cr$_2$Ge$_2$Se$_6$ and Cr$_2$Ge$_2$Te$_6$ by DFT calculations. The formation energy of Cr$_2$Ge$_2$Se$_6$ is given by  $E_{form} = E_{tot} - 2E_{Cr} - 2E_{Ge} - 6E_{Se}$, where $E_{tot}$ is the total energy of the 2D Cr$_2$Ge$_2$Se$_6$, and $E_{Cr}$, $E_{Ge}$ and $E_{Se}$ are the total energy per atom for the bulk chromium, germanium and selenium, respectively. We found that the formation energy of monolayer Cr$_2$Ge$_2$Se$_6$ is -1.13 eV, lower than the 0.92 eV of monolayer Cr$_2$Ge$_2$Te$_6$. It suggests that the 2D Cr$_2$Ge$_2$Se$_6$ should be more stable than 2D Cr$_2$Ge$_2$Te$_6$, the latter was realized in recent experiment \cite{Gong2017}.

 The electronic band structure of Cr$_2$Ge$_2$Se$_6$ is shown in Fig. \ref{F-2}(b). An indirect band gap of 0.748 eV is observed, and monolayer Cr$_2$Ge$_2$Se$_6$ is a semiconductor. Recently, it has been found that the change of the magnetization direction can alter the band structures in 2D CrI$_3$ \cite{nanoletterCrI3}. In our case, we calculate the band structure of Cr$_2$Ge$_2$Se$_6$ with the in-plane magnetization, and find that the band structure does not change dramatically. In addition, we uncover that a larger band gap of 1.38 eV is obtained by using the HSE06 hybrid functional, whereas the profiles of band structure do not change.

\section{Magnetic Properties}
\begin{figure}[tbp]
\includegraphics[width=8.5cm]{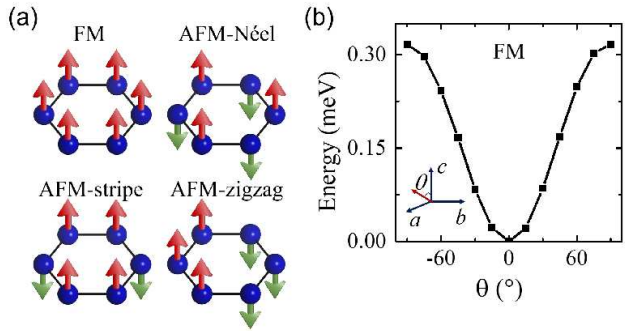}
\caption{(a) Four possible sates for two-dimensional Cr$_2$Ge$_2$Se$_6$: ferromagnetic (FM) state, antiferromagnetic (AFM) N\'{e}el state, AFM stripe state, and AFM zigzag state. Balls denote Cr atoms, and arrows denote spins. (b) Total energy as a function of angle $\theta$ for the FM state, obtained by the spin-polarized GGA + SOC + U calculations.}
\label{F-3}
\end{figure}

To study the magnetic ground state of 2D Cr$_2$Ge$_2$Se$_6$, we examine the possible states with paramagnetic, ferromagnetic and antiferromagnetic configurations. The calculations reveal that the energy of the paramagnetic state is 6 eV higher than the ferromagnetic and antiferromagnetic sates. Fig. \ref{F-3}(a) shows four possible magnetic sates of 2D Cr$_2$Ge$_2$Se$_6$: ferromagnetic (FM) state, antiferromagnetic (AFM) N\'{e}el state, AFM stripe state, and AFM zigzag state. The calculations show that the FM state is the most stable state, and the energy difference between the FM and the AFM configuration is more than 30 meV as listed in Table \ref{T-1}. Furthermore,  as shown in Fig. \ref{F-3}(b), the calculations show that the lowest energy of ferromagnetic state is obtained when the magnetization direction is perpendicular to the two-dimensional materials, with magnetic anisotropy energy of 0.32 meV per unit cell of Cr$_2$Ge$_2$Se$_6$. As listed in Table \ref{T-1}, the calculation shows that for 2D Cr$_2$Ge$_2$Se$_6$, the spin momentum of Cr atom is 3.4 $\mu_B$, and the occupation number of Cr 3$d$ orbitals is 3.976.

\begin{figure}[tbp]
\includegraphics[width=8.5cm]{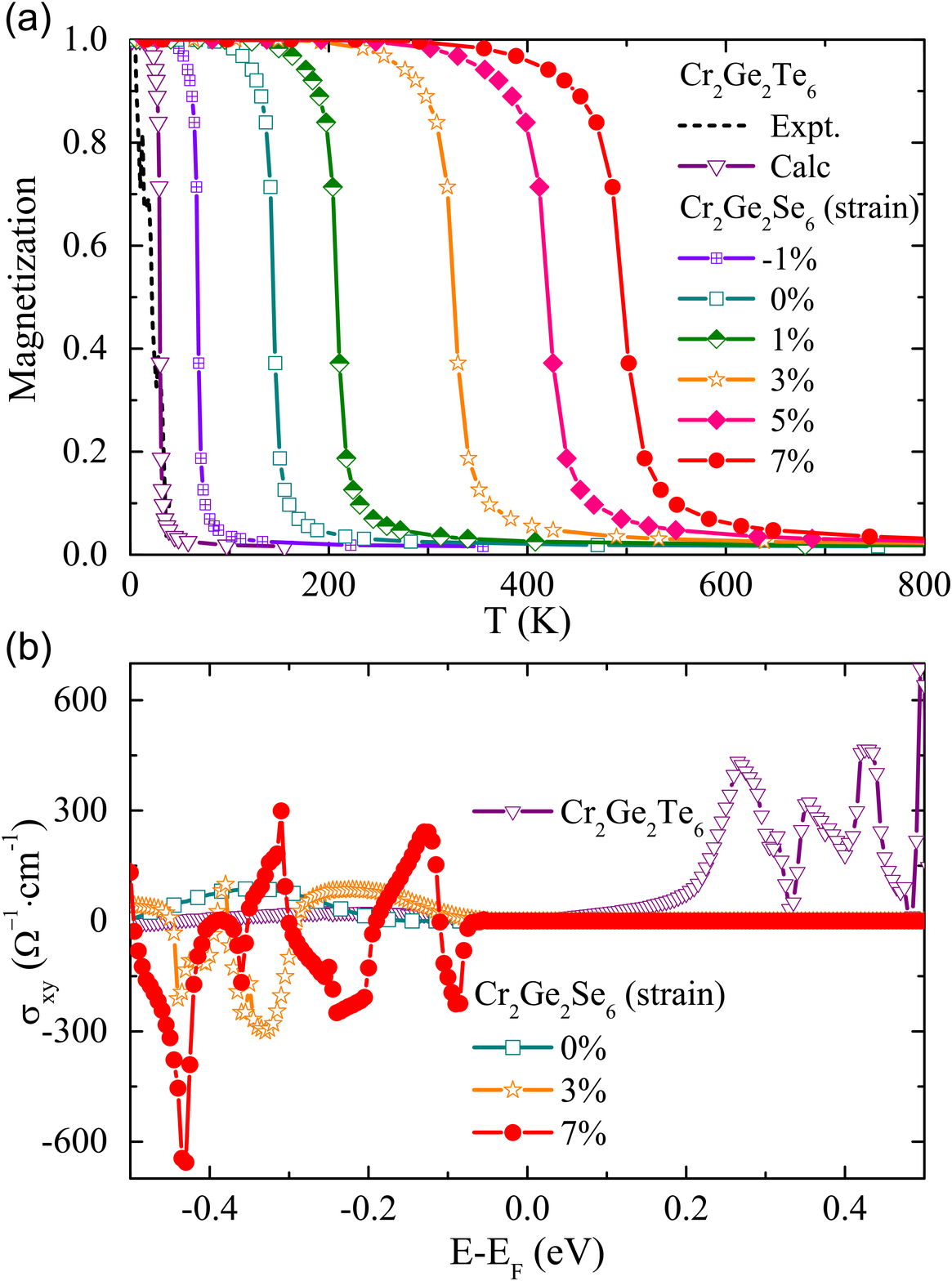}
\caption{For two-dimensional Cr$_2$Ge$_2$Te$_6$ and Cr$_2$Ge$_2$Se$_6$ with different strains, (a) the normalized magnetization as a function of temperature, and (b) the anomalous Hall conductivity as a function of energy. The experimental result of Cr$_2$Ge$_2$Te$_6$ is taken from Ref. \cite{Gong2017}. The calculation results are obtained by the DFT calculations and Monte Carlo simulations.}
\label{F-4}
\end{figure}

\begin{table}
\renewcommand\arraystretch{1.25}
\caption{For 2D semiconductor Cr$_2$Ge$_2$Te$_6$, Cr$_2$Ge$_2$Se$_6$, and Cr$_2$Ge$_2$Se$_6$ with 5\% strain, DFT results of total energy of ferromagnetic state E$^{\text{FM}}$, total energy of antiferromagnetic state E$^{\text{AFM}}$, Curie temperature $T$$_c$, 3$d$ orbital occupation number $n_d$, spin ($S$) and orbital ($L$) momentum of Cr atom, and bond length $d_{Cr-Te/Se}$ .}
\begin{tabular}{lccc}
\hline
\hline
 \multirow{2}{*}{DFT} &\multirow{2}{*}{Cr$_2$Ge$_2$Te$_6$}  & \multicolumn{2}{c}{Cr$_2$Ge$_2$Se$_6$}  \\
\cline{3-4}
 & & no strain   & 5\% strain \\
\hline
E$^{\text{FM}}$(eV)   & -89.1376  & -97.3460    & -96.5658   \\
 E$^{\text{AFM}}$(eV) & -89.1309   & -97.3135    & -96.4710   \\
E$^{\text{AFM}}$-E$^{\text{FM}}$(meV) & 6.7    & 32.5   & 98.8   \\
    $T$$_c$(K)   & 30       & 144     & 421  \\
$n_d$(Cr)      & 4.043    & 3.976   & 3.951 \\
S($\mu_B$)(Cr) & 3.586    & 3.404   & 3.487 \\
L($\mu_B$)(Cr)& -0.018  & -0.076 & -0.105 \\
$d_{Cr-Te/Se}$ ($\AA$) & 2.827 & 2.64 & 2.697 \\
\hline
\hline
\end{tabular}
\label{T-1}
\end{table}		

The Curie temperature $T$$_c$ can be estimated by the Monte Carlo simulation based on a 2D Ising model. The exchange coupling parameter is estimated as the total energy difference of ferromagnetic and antiferromagnetic states E$^{\text{AFM}}$-E$^{\text{FM}}$ as listed in Table \ref{T-1}, which was obtained by the DFT calculations. The obtained normalized magnetization as function of temperature is shown in Fig.\ref{F-4} (a). The experimental result of magnetization for Cr$_2$Ge$_2$Te$_6$ is taken from temperature-dependent Kerr rotation of bilayer Cr$_2$Ge$_2$Te$_6$ with $T$$_c$ = 28 K \cite{Gong2017}. The calculated Curie temperature is $T$$_c$ = 30 K for the monolayer Cr$_2$Ge$_2$Te$_6$, which is close to the experimental value. It is noted that for simplicity the Ising model is applied in our Monte Carlo simulation, while the Heisenberg model with magnetic anisotropy was suggested in experiment \cite{Gong2017}. In addition, the estimated Curie temperature $T$$_c$ could be even larger by using the mean field approximation \cite{Liu2018-mean-field}.
The calculated Curie temperature for monolayer  Cr$_2$Ge$_2$Se$_6$ is $T$$_c$ = 144 K, which is about 5 times higher than the $T$$_c$ = 30 K for monolayer Cr$_2$Ge$_2$Te$_6$ by the Monte Carlo simulation. More interestingly, the Curie temperature can be enhanced to $T$$_c$ = 326 K by applying tensile 3$\%$ strain, and $T$$_c$ = 421 K with 5$\%$ strain. We find that $T$$_c$ is decreased to 67 K with 1$\%$ compression strain, as shown in Fig. \ref{F-4}(a), and the system becomes antiferromagnetic when applying 2$\%$ compression strain. Our result predicts that monolayer Cr$_2$Ge$_2$Se$_6$ by applying a few percent tensile strain can be a promising candidate for room-temperature ferromagnetic semiconductor.

As the magnetization direction is out-of-plane with an easy axis along $c$ direction in 2D Cr$_2$Ge$_2$Te$_6$ and Cr$_2$Ge$_2$Se$_6$, it is interesting to study the anomalous Hall conductivity due to the Berry curvature of band structure. By the DFT calculations, the results are shown in Fig. \ref{F-4}(b). The magnitude of anomalous Hall conductivity $\sigma_{xy}$ for the p-type Cr$_2$Ge$_2$Se$_6$ and n-type Cr$_2$Ge$_2$Te$_6$ can be as large as 4 $\times$ 10$^2$ ($\Omega$ cm)$^{-1}$. This value is comparable to the $\sigma_{xy}$ in some ferromagnetic metals, such as $\sigma_{xy}$ = 7.5 $\times$ 10$^2$  in bcc Fe \cite{Yao2004,Wang2006}, and $\sigma_{xy}$ = 4.8 $\times$ 10$^2$ ($\Omega$ cm)$^{-1}$ in fcc Ni \cite{Wang2007} due to the Berry curvature of band structures. More importantly, the estimated $\sigma_{xy}$ in 2D magnetic semiconductors Cr$_2$Ge$_2$Te$_6$ and Cr$_2$Ge$_2$Se$_6$ is an order of magnitude larger than the $\sigma_{xy}$ of classic diluted magnetic semiconductor Ga(Mn,As) \cite{Jungwirth2003,Sinova2002}.

\begin{figure}[tbp]
\includegraphics[width=8.5cm]{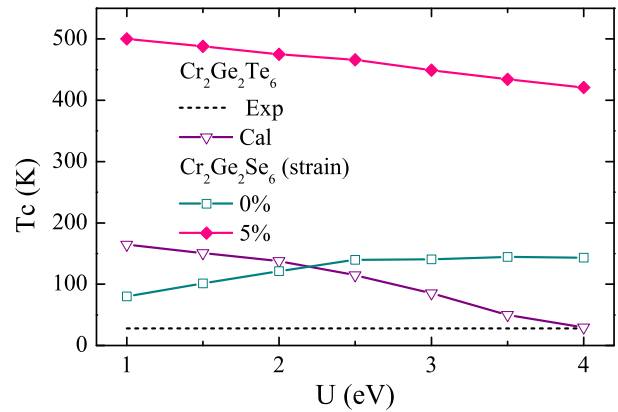}
\caption{  The Curie temperature $T_c$ as a function of parameter U for 2D Cr$_2$Ge$_2$Te$_6$, Cr$_2$Ge$_2$Se$_6$ without strain and with 5\% strain. The experimental result of Cr$_2$Ge$_2$Te$_6$ and the calculations are the same in Fig. \ref{F-4}(a), except with different values of U.}
\label{F-5}
\end{figure}

\section{Discussion}
To study the effect of the on-site Coulomb interaction U of 3$d$ orbitals of Cr, we calculate the Curie temperature $T_c$ as a function of parameter U for 2D Cr$_2$Ge$_2$Te$_6$, Cr$_2$Ge$_2$Se$_6$ without strain and with 5\% strain, as shown in Fig. \ref{F-5}. The experimental $T_c$ of 2D Cr$_2$Ge$_2$Te$_6$ is taken from Ref. \cite{Gong2017}. For 2D Cr$_2$Ge$_2$Te$_6$, the calculated $T_c$ is decreased with increasing U, and close to the experimental $T_c$ with U = 4 eV. In addition, we found that the magnetization direction of Cr$_2$Ge$_2$Te$_6$ becomes in-plane with U = 1 eV, which disagrees with experiment. The antiferromagnetic ground state is obtained for Cr$_2$Ge$_2$Te$_6$ with U = 5 eV, which is also against with the experiment. Our calculations suggest that the reasonable U for 2D Cr$_2$Ge$_2$Te$_6$ may be in the range of 2 $\sim$ 4 eV.

 For 2D Cr$_2$Ge$_2$Se$_6$ with 5\% strain, the calculated $T_c$ is always higher than room temperature with parameter U in the range of 1 $\sim$ 4 eV, as shown in Fig. \ref{F-5}. Thus, our prediction about the room temperature ferromagnetic semiconductor Cr$_2$Ge$_2$Se$_6$ with a few percent strain is robust with the values of parameter U.

\begin{table}
\renewcommand\arraystretch{1.25}
\caption{For 2D semiconductor Cr$_2$Ge$_2$Se$_6$ witout strain and with 5\% strain, DFT results of hopping matrix element $|V|$ and energy difference $|E_{p} - E_{d}|$ between 4$p$ orbitals of Se and 3$d$ orbitals of Cr.}
\begin{tabular}{c|ccccc}
\hline
\hline
\ strain & \multicolumn{5}{c}{hopping matrix element $|V|$ (eV)}  \\
\hline
       & $p_{z}$-$d_{z^2}$ & $p_{z}$-$d_{xz}$ & $p_{z}$-$d_{yz}$  & $p_{z}$-$d_{x^2-y^2}$  & $p_{z}$-$d_{xy}$\\
\cline{2-6}
0\%   & 0.257983 & 0.29527 & 0.380283 & 0.190041 & 0.445906  \\
5\%   & 0.258289 & 0.22747 & 0.306872 & 0.201524 & 0.416315   \\
\hline
\hline
     & \multicolumn{5}{c}{energy difference $|E_{p} - E_{d}|$ (eV)}  \\
\hline
       & $p_{z}$-$d_{z^2}$ & $p_{z}$-$d_{xz}$ & $p_{z}$-$d_{yz}$  & $p_{z}$-$d_{x^2-y^2}$  & $p_{z}$-$d_{xy}$\\
\cline{2-6}
0\%   & 1.415092 & 0.4458  & 0.45512 & 0.581666 & 0.581166  \\
5\%   & 1.552237 & 0.09143 & 0.09775 & 0.695055 & 0.693946   \\
\hline
\hline
\end{tabular}
\label{T-2}
\end{table}

 How to understand the enhancement of $T$$_c$ in 2D Cr$_2$Ge$_2$Se$_6$ by applying strain? According to the superexchange interaction \cite{Goodenough1955,Kanamori1960,Anderson1959}, the FM coupling is expected since the Cr-Se-Cr bond angle is close to 90 degree. The indirect FM coupling between Cr atoms is proportional to the direct AFM coupling between neighboring Cr and Se atom. The magnitude of this direct AFM coupling can be roughly estimated as $J = \frac{|V|^{2}}{|E_{p} - E_{d}|}$, where $|V|$ is the hopping matrix element between 4$p$ orbitals of Se and 3$d$ orbitals of Cr, and $|E_{p} - E_{d}|$ is the energy difference between 4$p$ orbitals of Se and 3$d$ orbitals of Cr. By DFT calculations, we can obtain these parameters for 2D Cr$_2$Ge$_2$Se$_6$ without strain and with 5\%  strain. The results of $|V|$ and $|E_{p} - E_{d}|$ are listed in Table \ref{T-2}. The results suggest that the strain has little effect on the hopping $|V|$, while the strain can make energy difference between $p_{z}$ orbital of Se and $d_{xz}$ and $d_{yz}$ orbitals of Cr decreased to one-fifth of the value without strain, causing a dramatical enhancement of the AFM coupling between Cr and Se atom. This leads to the enhanced FM coupling between Cr atoms, and makes $T_c$ beyond the room temperature.

\section {Summary}
By the DFT calculations we predict a stable 2D ferromagnetic semiconductor Cr$_2$Ge$_2$Se$_6$. We find the Curie temperature $T$$_c$ of Cr$_2$Ge$_2$Se$_6$ can be dramatically enhanced beyond room temperature by applying 3$\%$ strain, which is much higher than the $T$$_c$ = 28 K in 2D Cr$_2$Ge$_2$Te$_6$ in recent experiment. In addition, the anomalous Hall conductivity in 2D Cr$_2$Ge$_2$Se$_6$ and Cr$_2$Ge$_2$Te$_6$ is predicted to be an order of magnitude larger than that in diluted magnetic semiconductor Ga(Mn,As). Based on the superexchange interaction, the decreased energy difference between 4$p$ orbitals of Se and 3$d$ orbitals of Cr is found to be important to enhance the $T_c$ in 2D Cr$_2$Ge$_2$Se$_6$ by applying strain. Our finding highlights the microscopic mechanism to obtain the room temperature ferromagnetic semiconductors by strain.

\section* {Acknowledgments}
The authors acknowledge Q. B. Yan, Z. G. Zhu, and Z. C. Wang for many valuable discussions. BG is supported by NSFC(Grant No. Y81Z01A1A9), CAS (Grant No.Y929013EA2) and UCAS (Grant No.110200M208). GS is supported in part by the the National Key R\&D Program of China (Grant No. 2018FYA0305800), the Strategic Priority Research Program of CAS (Grant Nos.XDB28000000, XBD07010100), the NSFC (Grant No.11834014), and Beijing Municipal Science and Technology Commission (Grant No. Z118100004218001).


%

\end{document}